\documentstyle[multicol,aps,floats]{revtex}
\newcommand{\calD}{{\cal D}}
\newcommand{\calL}{{\cal L}}
\newcommand{\calP}{{\cal P}}
\newcommand{\calN}{{\cal N}}
\newcommand{\calH}{{\cal H}}
\newcommand{\tr}{{\rm tr}}
\newcommand{\psibar}[1]{\bar{\psi} ^{#1}}

\begin{document}
\draft                                        
\preprint{}                                   
\title{ Two-dimensional Dirac fermions with 
random axial-vector potential} 
\author{ Takahiro Fukui \cite{Email}, 
Hayato Emura, and Hiroki Yamada }
 \address{
Department of Mathematical Sciences,
Ibaraki University, Mito 310-8512, Japan }
\date{\today}
\maketitle
\begin{abstract}
A Dirac fermion model with random axial-vector potential is proposed.
At a special strength of randomness, the symmetry of the
action is enhanced, which is due to the gauge symmetry \`a la 
Nishimori. Some exact scaling exponents of single-particle 
Green functions are computed. The relationship with the 
XY gauge glass model is discussed.
\end{abstract}

\pacs{PACS: 72.15.Rn, 71.23.-k, 05.30.-d, 11.10.-z}

\begin{multicols}{2}

Random Dirac fermions have attracted much current interest 
in condensed matter physics.
They have actually intimate relationship 
with integer quantum Hall (IQH) transitions\cite{LFSG,MCW,Ber},
dirty $d$-wave superconductors\cite{DWV,CLD}, etc.
Critical theory of IQH transitions is believed to be a strong
coupling fixed point of the Dirac fermion with random mass,
vector and scalar potentials\cite{LFSG}, 
though the fixed point is still missing.
Near the zero energy quasiparticles of
disordered $d$-wave superconductors can be 
described by the Dirac fermion,
giving four kinds of novel universality classes\cite{Zir,AltZir}.
Including the conventional orthogonal, unitary and symplectic classes
and corresponding chiral classes, 
ten universality classes have been established
for disordered systems\cite{Zir}.

On the other hand, it has been reported recently that several 
other symmetry classes should exist\cite{RedGre,GLR,Iva}.
One example is realized\cite{GLR}
on the Nishimori-line \cite{Nis} 
of the random-bond Ising model\cite{RBI}.
The random-bond 
Ising model has O$(2n)$ symmetry\cite{DotDot,Sha}, 
belonging to the 
universality class $D$ \cite{CLD}
(the class of superconductors with
broken spin rotation and time reversal symmetry)
in Zirnbauer's classification \cite{Zir}.
Although the enhancement of the symmetry on the Nishimori-line 
has already been discussed in \cite{DouHar},
Gruzberg et. al. have explicitly shown that 
the symmetry is enhanced to O$(2n+1)$ 
due to the gauge symmetry \cite{GLR}.
This class is not included in Zirnbauer's classification.

Among these developments, 
the Dirac fermion with random vector potential only
is one of simplest but nontrivial models,
still providing hot topics.
It gives multifractal scaling exponents of local composite 
operators \cite{LFSG,MCW,Ber}, 
exact zero-energy wave function for any realization of 
disorder\cite{LFSG,MCW,CCFGM}, 
replica symmetry breaking\cite{CarDou,RyuHat}, etc.

In this paper, we study a Dirac fermion in two dimensions
including {\it random axial-vector potential}.
We show that the symmetry of the model is enhanced at 
a strong disorder strength, which is analogous to the 
Nishimori-line of statistical models.
Similarly to the conventional model with random vector potential,
this model can be solved exactly, giving us some exact scaling
dimensions.
It turns out that this model corresponds to the spin wave model for
the XY gauge glass model.

The system we will study is
\begin{eqnarray}
&&
\calH_A=\psibar{}h_A\psi, \quad
h_A=
\gamma_\mu \left(-i\partial_\mu-\sqrt{g}\gamma_5A_\mu\right) ,
\label{Ham}
\end{eqnarray}
where $\gamma_1$, $\gamma_2$, and $\gamma_5$ denote the three
Pauli matrices, and $A_\mu$ is a random axial-vector potential.
Being defined in 2D Euclidean space,
this Hamiltonian is non-hermitian due to $\gamma_5$ coupled with
the vector potential.
Therefore, the model (\ref{Ham}) may have a relationship with
Hatano-Nelson model \cite{HatNel}.
However, ensemble-average yields the square 
of the antihermitian term, 
and the Lagrangian becomes hermitian.
It should also be mentioned that without mass, the Hamiltonian 
has chiral symmetry
$\gamma_5h_A+h_A\gamma_5=0$, telling that the model belongs to
the class $A$III in Zirnbauer's classification. 

In the previous work \cite{LFSG,MCW,Ber}, 
the same model but with usual vector potential,
$h_V=\gamma_\mu\left(-i\partial_\mu-\sqrt{g}A_\mu\right)$
was studied. In the present case, 
the coupling with $\gamma_5$ may imply that
the right- and left-moving fermions have opposite charges.
In what follows, the present model (\ref{Ham}) is 
sometimes referred to as A-model, 
whereas the conventional one as V-model. 
The probability distribution of the axial-vector potential
is assumed to be of Gaussian type
\begin{eqnarray}
P[A]=\frac{1}{N_A}e^{-\frac{1}{2}\int d^2xA_\mu^2} ,
\label{ProDis}
\end{eqnarray}
where the normalization factor is given by
$N_A=\int\calD Ae^{-\frac{1}{2}\int d^2xA_\mu^2}$.
At a given Matsubara frequency $\omega$,
the partition function of the model with quenched disorder
is given by
\begin{eqnarray}
Z=\int\calD\psi\calD\psibar{} 
e^{-\int d^2x\psibar{}\left(h_A+i\omega\right)\psi}.
\label{Par}
\end{eqnarray}
The Lagrangian is invariant under the following
chiral gauge transformation,
\begin{eqnarray}
&&\psi\rightarrow e^{i\sqrt{g}\theta(x)\gamma_5}\psi,\quad
\psibar{}\rightarrow \psibar{} e^{i\sqrt{g}\theta(x)\gamma_5},
\nonumber\\
&&A_\mu\rightarrow A_\mu+\partial_\mu\theta(x) .
\end{eqnarray}
It is well-known that this transformation gives rise to a 
nontrivial Jacobian, but as we shall see later, 
it is irrelevant to our problem.
The $\omega$-term serves as a symmetry-breaking term.

Although it is possible to take quenched average of the 
partition function without resort to the replica trick,
we will first apply it in order to explain the enhancement 
of the symmetry at a special coupling constant.
Integration over disorder for replicated partition function
$Z^n$ yields an effective Lagrangian, 
$\calL_A=\psibar{a}
\left(-\gamma_\mu i\partial_\mu+i\omega\right)\psi^a
-\frac{g}{2}\left(\psibar{a}\gamma_5\gamma_\mu\psi^a\right)^2$,
where $a=1,\ldots,n$.
Due to the identity 
$\gamma_\mu\gamma_5=-i\epsilon_{\mu\nu}\gamma_\nu$,
we have 
$-\frac{g}{2}\left(\psibar{a}\gamma_5\gamma_\mu\psi^a\right)^2
=+\frac{g}{2}\left(\psibar{a}\gamma_\mu\psi^a\right)^2$.
Compared with the effective Lagrangian of the V-model,
$\calL_V=\psibar{a}
\left(-\gamma_\mu i\partial_\mu+i\omega\right)\psi^a
-\frac{g}{2}\left(\psibar{a}\gamma_\mu\psi^a\right)^2$,
the present A-model turns out to be equivalent to the V-model
but with negative coupling constant $g$.
Namely, the partition function $\overline{Z^n_A}(g)$ of 
the A-model with the coupling constant $g>0$ is equivalent
to that of the V-model $\overline{Z^n_V}(-g)$. 
Therefore, some properties, e.g., the scaling dimensions of 
some local operators for the A-model can be obtained 
directly from the 
V-model by the use of the {\it formal} replacement $g\rightarrow -g$,
as we shall see momentarily.

However, the A-model has some peculiar properties. 
One of them is the enhancement of the symmetry at 
a special coupling constant $g=\pi$. 
The Lagrangian in Eq. (\ref{Par}) has global
U($n$)$\times$ U($n$) symmetry for any $g$ when $\omega=0$. 
Explicitly, 
it is readily seen that for any realization of the axial-vector 
potential it is invariant under the transformation 
$\psi\rightarrow (U_1\calP_++U_2\calP_-)\psi$ and
$\psibar{}\rightarrow 
\psibar{}(U_1^\dagger\calP_-+U_2^\dagger\calP_+)$,
where $\calP_\pm=(1\pm\gamma_5)/2$ and
$U_i$ is a $n\times n$ unitary matrix which acts on the
replica space.
Since the Lagrangian is invariant under local chiral gauge
transformation, we have
\begin{eqnarray}
\overline{Z^n}&&=\int\calD\psi\calD\psibar{}\calD A
e^{-\frac{1}{2}\int d^2x A_\mu^2}
e^{-\int d^2x\calL}
\nonumber\\
&&=\frac{1}{V}\int\calD\psi\calD\psibar{}\calD A\calD\theta
e^{in\Gamma[\theta]}
e^{-\frac{1}{2}\int d^2x (A_\mu-\partial_\mu\theta)^2}
e^{-\int d^2x\calL}  ,
\nonumber\\
\end{eqnarray}
where 
$\calL=-\psibar{a}\gamma_\mu
\left(i\partial_\mu+\sqrt{g}\gamma_5A_\mu\right)\psi^a$,
$V$ is the ``volume'' of the gauge space $V=\int\calD\theta$, and
$\Gamma(\theta)$ 
is a Jacobian due to chiral gauge transformation \cite{fuj}.
This equation is based on 
the same argument as Nishimori applied to the random-bond
Ising model \cite{Nis}.
Since the Jacobian is multiplied by a factor $n$, it should vanish 
in the replica limit $n\rightarrow0$. 
This is also justified below in the direct computation of the exact
scaling dimensions of the single-particle Green functions.
Now using the fermionization of the gauge variable via
fermion-boson correspondence in 2D,
\begin{eqnarray}
\frac{1}{2}(\partial_\mu\theta)^2\leftrightarrow
-\psibar{0}\gamma_\mu i\partial_\mu\psi^0, \quad
\frac{i}{\sqrt{\pi}}\epsilon_{\mu\nu}\partial_\nu\theta
\leftrightarrow\psibar{0}\gamma_\mu\psi^0,
\label{Bos}
\end{eqnarray}
we have
\begin{eqnarray} 
\overline{Z^n}
&&
=\frac{1}{V}\int\calD\psi\calD\psibar{}\calD A
e^{-\frac{1}{2}\int d^2x A_\mu^2}
e^{-\int d^2x\calL} ,
\end{eqnarray}
where the Lagrangian now includes an additional 0th fermion
$\calL=
-\psibar{\alpha}\gamma_\mu i\partial_\mu\psi^\alpha
-\sqrt{g}(\psibar{a}\gamma_\mu\gamma_5\psi^a
+\sqrt{\frac{\pi}{g}}\psibar{0}\gamma_\mu\gamma_5\psi^0)A_\mu$.
The Greek superscript $\alpha$ runs from 0 to $n$, whereas
$a$ from 1 to $n$.
When $g=\pi$ (and $\omega=0$), the ensemble-averaged
Lagrangian can be written 
in a symmetric way together with the 0th fermion as 
$\calL=
-\psibar{\alpha}\gamma_\mu i\partial_\mu\psi^\alpha
+\frac{\pi}{2}
\left(\psibar{\alpha}\gamma_\mu\psi^\alpha\right)^2$,
telling that the symmetry is enhanced to 
U($n+1$)$\times$ U($n+1$). 
This enhancement of the symmetry for the Dirac fermion
with random {\it axial-vector potential}
is quite analogous to that for the Majorana 
or Dirac fermion with random {\it mass}
(the random-bond Ising model). 
Interestingly, such an enhancement does not occur in the case
with conventional {\it vector potential}.

What happens at this special coupling constant? For the
conventional model $h_V$, the exact scaling properties have
been obtained by various methods\cite{LFSG,MCW,Ber}. 
Since the present Hamiltonian 
$h_A$ is nonhermitian, we exactly solve the model by
taking the quenched average directly, without using 
the replica trick or the SUSY technique.
One of most interesting properties in disordered systems
is the multifractality of the scaling exponents.
The Dirac fermion with random vector potential is a 
typical example having such exponents.
By the use of the single-particle Green function,
\begin{eqnarray}
G_{ij}(x,y,i\omega)=\langle\psi_i(x)\psibar{}_j(y)\rangle,
\end{eqnarray}
where 
\begin{eqnarray}
\langle O\rangle=\frac{1}{Z}\int\calD\psi\calD\psibar{}
Oe^{-\int d^2x\calL} ,
\label{ExpVal}
\end{eqnarray}
ensemble-averaged density of state (DOS) 
and a field-theoretical analogue of
the inverse participation ratios (IPR)
$P^{(k)}(E)=\overline{\sum_n|\Psi_n(x)|^{2k}\delta(E-E_n)}/
\overline{\sum_n|\Psi_n(x)|^{2}\delta(E-E_n)}$
can be computed as \cite{Weg}
\begin{eqnarray}
&&
\rho(E)=-\frac{1}{\pi}\lim_{\omega\rightarrow0}
{\rm Im}~\tr\overline{G(x,x,i\omega-E)} ,
\nonumber\\
&&
P^{(k)}(E)\rho(E)=\frac{1}{C_k}\lim_{\omega\rightarrow0}
\omega^{k-1}\overline{[{\rm Im}~\tr G(x,x,E-i\omega)]^k} ,
\label{DOSIPR}
\end{eqnarray}
where $\Psi_n(x)$ is a two component spinor wave function of 
$n$th eigenstate, $\tr$ is a trace for spinor indices, and
$C_k=\pi(2k-3)!!/(2n-2)!!$.
It is useful to introduce the chiral basis,
$\psi_\pm=\calP_\pm\psi$ and $\psi^\dagger_\pm=\psibar{}\calP_\mp$,
which was actually used by Mudry et. al.\cite{MCW}.
In what follows, we will take similar notations to this reference.
Since $\psibar{}\psi=\psi^\dagger_+\psi_-+\psi^\dagger_-\psi_+$,
the $k$th power of the Green function can be written as
\begin{eqnarray}
\overline{\left[G(x,x,E-i\omega)\right]^k}
\propto\Big\langle\prod_{a=1}^k
\left[\psi^{\dagger a}_+\psi_-^a(x)
+\psi^{\dagger a}_-\psi_+^a(x)\right]\Big\rangle .
\end{eqnarray}
Here we have introduced species $a$ to calculate the $k$th
power of the Green function, although we will not use 
the replica trick.
Expanding the product, we have various operators with various
scaling dimensions. Most dominant operators depends on whether
$k$ is even or odd. Define 
\begin{eqnarray}
O^{(k)}(x)=
\prod_{a=1}^{[k/2]}\psi^{\dagger2a-1}_+\psi_-^{2a-1}
\psi^{\dagger2a}_-\psi_+^{2a}(x),
\end{eqnarray}
then an example of most dominant operator is 
$O_e^{(k)}=O^{(k)}$ for even $k$ and 
$O_o^{(k)}=O^{(k)}\psi^{\dagger k}_+\psi_-^k$ for odd $k$.

In the chiral basis the Lagrangian is converted into
\begin{eqnarray}
\calL=
\psi^{\dagger a}_+(2i\partial_{\bar{z}}-\sqrt{g}A_+)\psi_+^a
+\psi^{\dagger a}_-(2i\partial_z+\sqrt{g}A_-)\psi_-^a,
\end{eqnarray}
where
$z=x+iy$, $\bar{z}=x-iy$, $A_\pm=A_x\pm iA_y$,
and $a=1,2,\cdots,k$.
The different sign for $A_\pm$ tells that the right- and 
left-movers have opposite charges, as mentioned previously.
In 2D the gauge fields can be decomposed into two independent
components 
$A_\mu=\epsilon_{\mu\nu}\partial_\mu\eta+\partial_\mu\xi$,
or in the chiral basis,
$A_+=-2i\partial_{\bar{z}}(\eta+i\xi)$, and
$A_-=2i\partial_z(\eta-i\xi)$.
The probability distribution (\ref{ProDis}) is written now 
in terms of these fields as
\begin{eqnarray}
P[\xi,\eta]=\frac{1}{N_\xi N_\eta}
e^{-\frac{1}{2}\int d^2x
\left[(\partial_\mu\xi)^2+(\partial_\mu\eta)^2\right]},
\label{ProDis2}
\end{eqnarray}
where $N_{\xi,\eta}$ is a normalization factor 
$N_\xi=\int\calD\xi e^{-\frac{1}{2}\int d^2x(\partial_\mu\xi)^2}$
and similar for $N_\eta$.
Via the gauge transformation
$\psi^a_\pm\rightarrow e^{\sqrt{g}(\eta\pm i\xi)}\psi^a_\pm$ and
$\psi^{\dagger a}_\pm\rightarrow\psi^{\dagger a}_\pm 
e^{-\sqrt{g}(\eta\pm i\xi)}$,
the axial-vector potential completely disappears in the Lagrangian,
\begin{eqnarray}
\calL\rightarrow
\psi^{\dagger a}_+2i\partial_{\bar{z}}\psi_+^a
+\psi^{\dagger a}_-2i\partial_z\psi_-^a .
\label{FreLag}
\end{eqnarray}
The chiral gauge transformation yields a nontrivial Jacobian
$\calD\psi\calD\psibar{}\rightarrow
\calD\psi\calD\psibar{}e^{ik\Gamma(\xi)}$ \cite{fuj}.
However, this factor appears in the denominator $Z$ as well as
in the numerator in Eq. (\ref{ExpVal}), and hence cancels out.
Finally, $O_e^{(k)}$ is gauge-invariant while $O_o^{(k)}$ is not.
Actually, they obey the transformation laws
\begin{eqnarray}
O_e^{(k)}\rightarrow O_e^{(k)}, \quad
O_o^{(k)}\rightarrow e^{-2i\sqrt{g}\xi}O_o^{(k)} .
\end{eqnarray}
By the use of the free action (\ref{FreLag}) after the 
gauge transformation, 
the correlation function of $O_{e,o}^{(k)}$ can be evaluated
as follows.
The two point correlators of the free fermi fields are 
$\langle\psi_+^a(z)\psi^{\dagger a'}_+(0)
\rangle\sim \delta^{aa'}z^{-1}$
and 
$\langle\psi_-^a(z)\psi^{\dagger a'}_-(0)
\rangle\sim \delta^{aa'}\bar{z}^{-1}$.
Therefore, we have
\begin{eqnarray}
&&
\langle O_e(x)O_e^\dagger(0)\rangle=|x|^{-2k},
\nonumber\\
&&
\langle O_o(x)O_o^\dagger(0)\rangle
=e^{-2i\sqrt{g}\xi(x)}e^{2i\sqrt{g}\xi(0)}|x|^{-2k}.
\end{eqnarray}
Note that these do not depend on $\eta$ and therefore 
the change of the scaling dimension due to disorder
is involved with $\xi$ only.
The average-over $\xi$ is easily taken and 
the scaling dimension of $O_{e,o}^{(k)}$ finally reads
$\Delta_k=k$ for even $k$ whereas 
$\Delta_k=k+\frac{g}{\pi}$ for odd $k$, and this is itself
nothing but the dominant scaling dimension of the $k$th power 
of the single-particle Green function $\overline{(\tr G)^k}$. 
Some comments may be in order. In the conventional
V-model the random field $\eta$ plays a role in the change of the
dimensions due to disorder. 
In the present case, the role of $\eta$ and $\xi$ is exchanged
by $\gamma_5$ in Eq. (\ref{Ham}),
and $\xi$ causes the change of the scaling dimensions. 
Furthermore, the field $\eta$ gives in general 
negative scaling dimensions for the V-model, whereas
$\xi$ gives {\it positive} dimensions for the present model.

Using the dominant scaling dimensions of the Green functions
obtained so far,
we can calculate the scaling properties of the DOS and IPR.
Since $\omega$ couples with $\psibar{}\psi$ in the Lagrangian,
the dimension of $\omega$ reads 
$z=1-\frac{g}{\pi}$ \cite{LFSG,MCW}.
Therefore, we expect
$\rho(\omega)\sim\omega^{(2-z)/z}$. 
This is just the same formula for the V-model but with negative $g$. 
It is also readily seen that 
the random axial-vector potential is a marginal perturbation
and the theory moves along a critical line.
However, starting from the free fermion point, 
the A-model moves, as $g$ increses, to the opposite direction 
to which the V-model moves.
Therefore, the line on which the A-model lies is not reached
by the conventional random vector potential model.
In this sense, the present model is complementary to
the full critical line of U(1)$\times$U(1) symmetry.
The dimension of $\rho$ is an increasing function of $g$, so that 
the DOS is suppressed around the zero energy 
if the random axial-vector potential is switched on.
At the special point $g=\pi$ where the symmetry is enhanced,
the exponent becomes infinity.
Therefore, it is likely that the theory is well-defined only for
$g<\pi$, and the enhancement of the symmetry is a signal that
the theory reached at a singularity.

The scaling behavior of the IPR is obtained in a similar way:
$P^{(k)}\sim\omega^{\tau_k^*/z}$, where
$\tau_k^*=-2$ for even $k$ and $\tau_k^*=0$ for odd $k$.
As stressed by Mudry et. al.\cite{MCW}, the IPR defined by
Eq. (\ref{DOSIPR}) is not necessarily coincides with that of 
the original definition
$P^{(k)}=\overline{|\Psi_n(x)/\calN|^{2k}}$, where
the normalization $\calN$ is defined by 
$\calN^2=\int d^2x|\Psi_0(x)|^2$.
Actually, the zero energy wave function can be obtained exactly as
\begin{eqnarray}
\Psi_0(x)=e^{\eta(x)-i\xi(x)\gamma_5}\varphi_0,
\label{ZerEne}
\end{eqnarray}
for the present A-model, where $\varphi_0$ is a constant spinor,
while 
$\Psi_0(x)=e^{\eta(x)\gamma_5-i\xi(x)}\varphi_0$ 
for the conventional V-model.
The IPR of the latter wave function has been studied by various
methods, e.g., replica method\cite{LFSG}, SUSY method\cite{MCW},
the method using the equivalence to the 
random energy model\cite{CCFGM}, etc.
They have all given the same scaling exponent for small $g$.
Eq. (\ref{ZerEne}) for the present model also gives 
$|\Psi_0(x)|^2\propto e^{2\eta(x)}$. 
Since the present A-model has the same
probability distribution (\ref{ProDis2})
as the conventional V-model has, we expect that the 
the A-model should have the same scaling dimension
$\tau_k=(2-\frac{g}{\pi}k)(k-1)$
as the V-model has, which is quite 
different from the conjectured $\tau_k^*$ based on (\ref{DOSIPR}).

So far we have studied some peculiar properties of the 
Dirac fermion with random axial-vector potential.
As already mentioned, 
the Hamiltonian is nonhermitian, because it is defined in 
the Euclidean space. 
Then, is this model unrealistic?
To address the question,
we next examine the bosonized form of the 
A-model, as has been done for the V-model by Bernard\cite{Ber}. 
By the use of the correspondence (\ref{Bos}),
the Hamiltonian (\ref{Ham}) is converted into
\begin{eqnarray}
\calH=\frac{1}{2}(\partial_\mu\phi)^2
+\sqrt{\frac{g}{\pi}}\partial_\mu\phi A_\mu .
\label{SpiWav}
\end{eqnarray}
This is a low-temperature effective Hamiltonian of the 
XY gauge glass model 
\cite{Nis,GauGla,OzeNis} whose Hamiltonian is defined on a
2D square lattice,
\begin{eqnarray}
-\beta H=K\sum_{\langle i,j\rangle}\cos(\phi_i-\phi_j+\chi_{ij}),
\label{HamXY}
\end{eqnarray}
where $\chi_{ij}$ is a random gauge variable 
with a probability distribution
\begin{eqnarray}
P[\chi]\propto e^{K_p\sum_{\langle i,j\rangle}\cos\chi_{ij}}.
\label{ProDisXY}
\end{eqnarray}
Provided that $K$ and $K_p$ are large,
we actually reach the Hamiltonian (\ref{SpiWav}) 
with $g=\frac{\pi K}{K_p}$ 
and the probability distribution (\ref{ProDis}) as well by
expanding the cosine terms up to second order, 
defining $\phi_i-\phi_j\sim a_0\partial_\mu\phi(x)$ and
$\chi_{ij}\sim a_0A_\mu(x)$ with the lattice constant $a_0$,
and rescaling the fields $\phi\rightarrow\phi/\sqrt{K}$
and $A_\mu\rightarrow A_\mu/\sqrt{K_p}$.
This is the so-called spin-wave approximation.
The spin-wave Hamiltonian (\ref{SpiWav}) is expected to be unstable
against vortex excitations if $K$ or $K_p$ becomes small.

On the other hand, it has been shown that the model (\ref{HamXY})
with (\ref{ProDisXY})
has a Nishimori-line $K_p=K$, on which some exact results can be
obtained in a similar way 
as the random-bond Ising model\cite{Nis,OzeNis}. 
The equivalence of the condition $K_p=K$ to $g=\pi$ 
implies that the symmetry enhancement of the Dirac fermion
with random axial-vector potential reflects the gauge symmetry on
the Nishimori-line of the gauge glass model. 
As to the symmetry of the model, it is difficult to read
its enhancement from the spin-wave Hamiltonian (\ref{SpiWav}).
Even if we apply the replica method, 
the symmetry always remains U(1).
This reminds us of the case of the random-bond Ising model:
We only come accross the continuous O($n$) symmetry
when the model has been described by the Majorana fermion 
via the Jordan-Wigner transformation. 
In the present case, the enhanced symmetry also becomes manifest 
only in the fermion description.
However, the Lagrangian discussed so far
is just for the spin-wave part of the gauge glass model and 
it is quite necessary to include vortices 
to fully describe the lattice model (\ref{HamXY}).
Then, such theories should be well-defined field theories 
with manifest enhanced symmetry on the Nishimori-line.
It is an interesting issue to derive and study 
such field theories.

Finally, let us discuss the symmetry for the dual theory
of the gauge glass model (\ref{HamXY})\cite{JKKN}.
The periodicity of the Hamiltonian can be described by
$e^{K\cos(\phi_i-\phi_j+\chi_{ij})}
\sim \sum_{m=-\infty}^\infty e^K
e^{-\frac{K}{2}(\phi_i-\phi_j+\chi_{ij}-2\pi m)}$.
By the use of the Poisson summation formula\cite{JKKN}, we reach 
the following Hamiltonian,
$
\calH_d=\frac{1}{2K}(\partial_\mu\theta)^2+
i\epsilon_{\mu\nu}A_\mu\partial_\nu\theta-2y\cos2\pi\theta,
$
where $\theta$ is a dual field and $y$ is a fugacity.
This is the sine-Gordon model coupled with  random 
{\it vector potential} $A_\mu$.
Therefore, it turns out that there is a duality between 
the random axial-vector potential and conventional vector
potential.
The former model yields a symmetry enhancement whereas the latter
does not. 
Recently, the replica symmetry-breaking of the V-model
has been suggested\cite{CarDou,RyuHat},
which might 
be understood via this duality relation.

In summary, we have studied a Dirac fermion with random 
axial-vector potential. 
This model has an enhanced symmetry at a special
strength of randomness, which reminds us of the Nishimori-line
of the statistical models.
Indeed, the model is equivalent to 
the spin wave Hamiltonian of the XY gauge glass model.
It turns out that 
this model moves along a critical line for increasing $g$
up to $\pi$, and some exact scaling exponents have been obtained.


One of the authors (TF) 
would like to thank H. Suzuki for valuable discussions.
This work is supported in part by the Yamada Science Foundation.


\end{multicols}

\end{document}